\documentclass[twocolumn,eqsecnum,showpacs,aps,epsfig,floats]{revtex4}

\usepackage{graphicx}  
\usepackage{subeqnar}  

\begin{document}

\draft \preprint{cond-mat/123456}


\title{Finite lattice size effect in the ground state phase diagram of
quasi-two-dimensional magnetic dipolar dots array with
perpendicular anisotropy}

\author{R. H. He$^{1}$, X. F. Jin$^{1}$}

\email{xfjin@fudan.ac.cn}

\address{$^1$Applied Surface Physics Laboratory, Physics Department, Fudan University, Shanghai 200433, China}


\date{\today}

\begin{abstract}

A prototype Hamiltonian for the generic patterned magnetic
structures, of dipolar interaction with perpendicular anisotropy,
is investigated within the finite-size framework by
Landau-Lifshift-Gilbert classical spin dynamics. Modifications on
the ground state phase diagram are discussed with an emphasis on
the disappearance of continuous degeneracy in the ground state of
in-plane phase due to the finite lattice size effect. The
symmetry-governed ground state evolution upon the lattice size
increase provides a critical insight into the systematic
transition to the infinite extreme.

\end{abstract}

 \vskip2pc
\narrowtext

\maketitle

Nanoscale magnetism of patterned magnetic structures (PDS) has
aroused a great deal of research interest due to its potential
technological applications\cite{NanoPattern_1} in high-density
magnetic storage media and spintronic devices such as magnetic
random access memory. Recent lithographic technologies have
rendered possible the design of various geometry of the
quasi-two-dimensional(2D) uniform array composed of identical
elements with a well-defined composition, shape and size in
sub-micrometer scale\cite{NanoPattern_2} and hence the control of
magnetic properties of the system. Each small-size dot, made up of
a large number of spins which interact ferromagnetically through
the intradot exchange interaction, tends to be kept in a
single-domain acting as a giant spin in response to the exerted
magnetic field\cite{NanoPattern_2}. As a contrast, the interdot
exchange interaction term is completely precluded from the generic
Hamiltonian in describing such an interacting dipole
system\cite{NanoPattern_3} because of the large interdot spacing,
\begin{equation}\label{eq:Hamiltonian}
    H_{int}=-D\sum_{i}S_z^2+U_{dipole} ,
\end{equation}
\begin{equation}\label{eq:DipolarTerm}
U_{dipole}=\frac{1}{2}\sum_{i,j}\Omega[\frac{1}{r_{ij}^3}(\vec{S}_i\cdot\vec{S}_j)
    -\frac{3}{r_{ij}^5}(\vec{r}_{ij}\cdot\vec{S}_i)(\vec{r}_{ij}\cdot\vec{S}_j)] ,
\end{equation}
where $D$ represents the on-site effective anisotropy strength,
which is the joint contributions by magnetocrystalline anisotropy
and shape anisotropy resulted from intradot dipolar coupling,
$U_{dipole}$ the interdot dipolar
interaction\cite{DipoleTermNote}, $\vec{S}_{i(j)}$ is the giant
spin at site $i(j)$, equal to the total moment of spins inside,
$\vec{r}_{ij}$ the vector connecting the two sites.

In recent years, driven by the efforts in resolving complex
micromagnetic mechanisms for, such as spin reorientation
transition\cite{DipolarPRL} and anti-ferromagnetic domain
nucleation\cite{DengAFMdomain}, as well as the growing extensive
interest in understanding the magnetism-related problems found in
various kinds of novel material whose interspin spacing is
relatively large in the atomic scale, such as high-spin
molecular\cite{HighSpin,SpinIce} and some high-temperature
superconductors with magnetic irons forming a quasi-2D
plane\cite{DDsquIn}, considerable theoretical attentions and
experimental efforts are put into the understanding of the
dominating dipolar effect involved in the prototype Hamiltonian.

Theoretically, however, as the uniqueness of PDS, the finite-size
nature, is seldom emphasized. Most analytical works based on
infinite dipole sums\cite{DipoleSum} as well as numerical works
using a period boundary condition
(PBC)\cite{DipolarPRL,DDsquIn,DDtriIn_1_2}(\textbf{\emph{more to
be cited here}}) do not practically apply to the PDS. Previous
results on the phase diagram of the easy-axis dipolar Hamiltonian
are expected to be adjusted, in the framework where a realistic
truncation on the dipole sums or the free boundary condition (FBC)
is used, before a direct comparison of with experiments on PDS can
be made\cite{DipolarPRL,DDsquIn}. On the other hand,
experimentally, efforts spent on the PDS are expected to be
rewarding in that they provide a rather handy way in finetuning
the relative strength of different interactions by only changing
the definable geometrical parameters of the system while keeping
the matrix material unchanged, and hence facilitates an easy probe
by the mature spacially-averaged measurements or
spacially-resolved imaging techniques\cite{NanoPattern_3} into a
wide range of phase diagram of the system, which in turn serves
the general understanding of the dipolar physics as long as the
precursory knowledge in the role of finite size is available.

In this paper, we attempt to build up the missing link in between
by presenting a systematic size-dependent evolution of the ground
state phase diagram of the easy-axis dipolar Hamiltonian. In
contrast to the robustness of the out-of-plane (OOP) phase, the
in-plane (IP) phase exhibits a pronounced modification under the
finite-size influence. The difference of the detailed dynamics of
the evolution as function of lateral size ($L$) for lattices with
an even and odd number $L$, is explained from the symmetry point
of view. The even-odd symmetry difference as well as the in-plane
anisotropic content of dipolar interaction on 2D lattice tend to
be concealed by its long-range nature upon $L\rightarrow\infty$,
as suggested by the extreme picture finally given.

\begin{figure}[t!]
\vspace*{-0.5cm} \hspace*{-0.5cm}
\centerline{\includegraphics[width=3.5in]{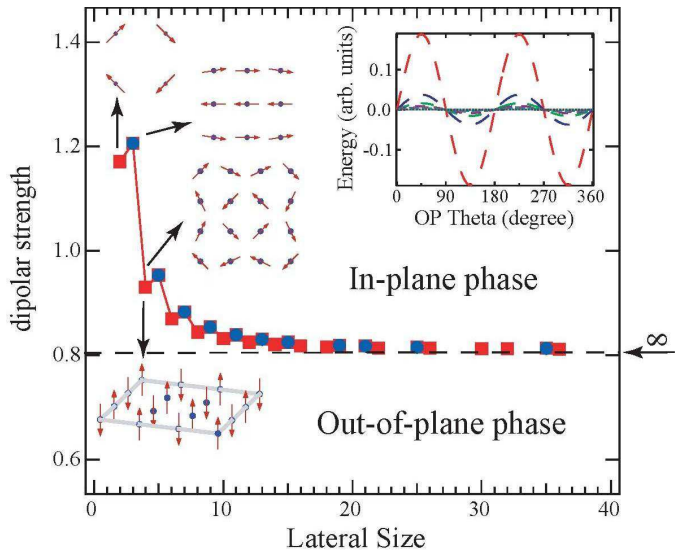}}
\caption[Finite-size phase diagram] {(color) The finite-size phase
diagram of the easy-axis dipolar Hamiltonian. The zigzag boundary
separates the in-plane phase (the upper half) from the
out-of-plane phase (the lower half). Insets in the in-plane phase:
the ground state spin configurations for $L= 2, 3, 4$,
respectively, and the OP theta-dependent in-plane AFM state energy
for $L=2, 4, 6, 8, 10, 14, 32$ (dash lines with decreasing length)
and all odd $L$'s (grey based line); the inset in the out-of-plane
phase: the out-of-plane AFM state spin configuration (3D) at $L=
4$.} \label{PhaseDiagram}
\end{figure}

Landau-Lifshift-Gilbert classical spin dynamics is strictly
followed to investigate the physical ground state and dynamical
properties of the system under zero field\cite{DengAFMdomain}. For
clarity, the module of spin vector, the anisotropy strength, the
gyromagnetic constant (as the unit reference for time and
effective field strength), and damping coefficient are set to
unity unless specified otherwise. Predictor-corrector method with
Runge-Kutta initialization is used to maintain a high accuracy of
the numerical integration of equation of motion\cite{ODE}, which
is required for the realistic comparison between our results with
experiments as well as true dynamics starting from a given initial
spin configuration. The investigation of finite lattice size
effect demands a differentiation on the free boundary condition
(FBC) and periodic boundary condition (PBC). Different from the
only use of PBC in MC simulation which always serves to approach
the infinite system with a well-defined temperature in the
statistical sense, our spin dynamics simulations are carried out
using FBC for the finite-size lattices and PBC for the infinite
one.

In Fig. \ref{PhaseDiagram}, we recap the the phase diagram of the
easy-axis dipolar Hamiltonian with its finite-size modifications.
For the infinite lattice, an out-of-plane (OOP) phase lies in the
small dipolar strength ($DD$) regime, whose ground state is
characterized by an OOP order parameter,
$\vec{M}_{z}=\frac{1}{N}\sum_{i}(-1)^{m+n} S^z$ (OOP-AFM), where
$m$ and $n$ are row and column index for the square lattice,
respectively; the planar nature of the 2D dipolar Hamiltonian
recovers by yielding an in-plane (IP) phase upon DD increase, with
its continuously degenerate ground state described by the IP order
parameter (OP) $\vec{M}_{xy}=\frac{1}{N}\sum_{i}[(-1)^n
S^x\hat{x}+(-1)^m S^y\hat{y}]$ (IP-AFM, $|\vec{M}_{xy}|$ the OP
module $|OP|$, $\arctan(\frac{M_y}{M_x})$ the OP theta). In the
OOP phase, when lattice size reduces to finite, the OOP-AFM state
(inset of Fig. \ref{PhaseDiagram}) remains stable regardless of
any specific lateral size ($L$). Exhaustive simulations (up to $L=
128$, 100~1000 per $L$, $DD= 0.5$) starting from random initial
configurations show this state has the lowest energy. The
robustness of the out-of-plane ground state points to the
persistence of 2D AFM Ising nature (due to the disappearance of
the second term in dipolar interaction) upon finite truncation in
the dipole sum\cite{DipoleSum}. As a contrast, a partial recovery
of the effective in-plane anisotropy at the cost of the long-range
nature of dipolar Hamiltonian removes largely the ground state
degeneracy in the in-plane phase. As shown in the inset of Fig.
\ref{PhaseDiagram}, for the even $L$'s, the IP-AFM states are no
longer continuously degenerate, whose energies exhibit a sine
distribution as a function of the OP theta with a rapidly
decreasing amplitude as $L$ increases; for all odd $L$'s,
interestingly, there is no difference in energy for these IP-AFM
states.

Notably, as for the IP-AFM states, the OP symmetry is different
for the even- and odd-$L$ lattices, which can be investigated
based on symmetry operations on the spin lattice. The axial OP
symmetry for the even-$L$ lattice is $\pm45^{\circ}$ while
$\pm45^{\circ}$ as well as $0^{\circ}$ and $90^{\circ}$ for the
odd-L lattice\cite{OPsymmetryNote}. Bearing this in mind, we
conduct a further examination on the stability of the IP-AFM
states at various $L$'s by independent spin dynamics simulations
taking them as the initial states and, similar as in the OOP phase
case, the ground state at each $L$ is confirmed by simulations
starting from random initial spin configurations. As illustrated
in the insets of Fig. \ref{PhaseDiagram}), the only ground state
of $L= 2$ is the IP-AFM state with the $135^{\circ}$ OP theta
(denoted as OP135, and similarly hereafter), which can be regarded
as a fully boundary-distorted (BD) spin configuration. When $L>
2$, the boundary effect is weakened as the number ratio between
the inside spins and the boundary spins increases. As seen in the
insets, the ground states of $L= 3, 4$ have clearly hybrid
compositions, whose central regions basically maintain the
original spin alignments while the peripheral spins tends to align
along the borders, and are denoted as BD-OP0 and BD-OP135,
respectively.

\begin{figure}[b!]
\centerline{\includegraphics[width=3.5in]{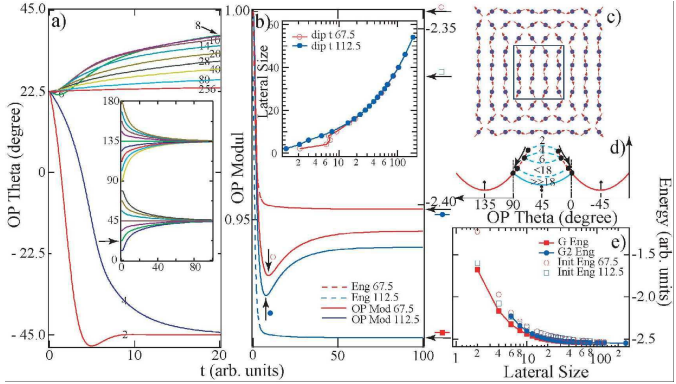}}
\caption[Even-L dynamics] {(color) Spin dynamics for the even-$L$
lattices. (a) $L$-dependent evolution of the OP22.5 state ($L$
indicated on curves). Inset: typical evolutions of various OP
theta states at intermediate $L$ ($L= 10$). (b) typical evolutions
in unit energy and $|OP|$ of OP67.5 and OP112.5 states at
intermediate $L$ ($L= 10$). The $L$-dependent relaxation time
positions (semilog) of the dips in the two $|OP|$ curves are
summarized in the inset. (c) the spin configuration ($L= 10$)
corresponding to the dip in the $|OP|$ curve of OP67.5, namely
BD-OP67.5 in the text. Note that spins in the framed region are
basically free from the boundary distortion. (d) schematics
showing the evolution of 3D free energy surface in the transverse
view. Numbers are $L$ values and arrows indicate the relative
motion of energy surface. (e) the $L$-dependent energies of the
ground state BD-OP135, the metastable state BD-OP45 and the
initial states OP67.5 and OP112.5 (semilog).} \label{EvenL}
\end{figure}

A prominent feature of the $L$-dependent phase boundary, which is
determined from the comparison in energy between the OOP and IP
ground states as a function of $DD$ for each $L$, is its zigzag
(oscillatory) decrease asymptotically to the value of infinite
lattice. The shrinkage of the out-of-plane phase upon $L$ increase
is the result of the long-range nature of the dipolar interaction
which favors in-plane magnetization; the non-monotonous behavior,
one of the characteristics of finite size effects frequently
encountered in other nanosciences\cite{Nanoscience}, not only
suggests the AFM nature of the dipolar ground state, but also
contains critical information about the dynamical evolution of the
system as it extends to infinite by following its unique OP
symmetry different for the even-$L$ case and the odd one.

In order to have a quantitative insight into the dynamical
evolution, we continue to use the OP expression on states during
the relaxation from the given initial states ($|OP|= 1$) though
their effective $|OP|$'s are expected to reduce dependent on the
OP inhomogeneity. A prototype of the $|OP|$ evolution is shown in
Fig. \ref{EvenL}(b), which is characterized by a preceding rapid
drop and a slow saturation. The boundary relaxation proceeds with
a major minimization on the total energy and ends up with a BD
state of the lowest $|OP|$, whose spin configuration is
exemplified in Fig. \ref{EvenL}(c) for OP67.5. The subsequent
relaxation mainly involves the rotation of OP to reach finally a
certain meta-stable state. The whole physical path is illustrated
in the inset of Fig. \ref{EvenL}(a) for initial states with
different OP theta values ($L= 10$). Notably, besides the ground
state BD-OP135, a meta-stable state, BD-OP45, forms at $L= 6$ by
attracting the initial states with OP theta in its proximity. For
example, in Fig. \ref{EvenL}(a), the OP22.5 state, which is
attracted to the BD-OP315 at $L= 2, 4$, experiences a pronounced
transformation in its physical path leading to a different final
state, BD-OP45, at $L > 6$.

\begin{figure}[t!]
\vspace*{-0.5cm} 
\centerline{\includegraphics[width=3.5in]{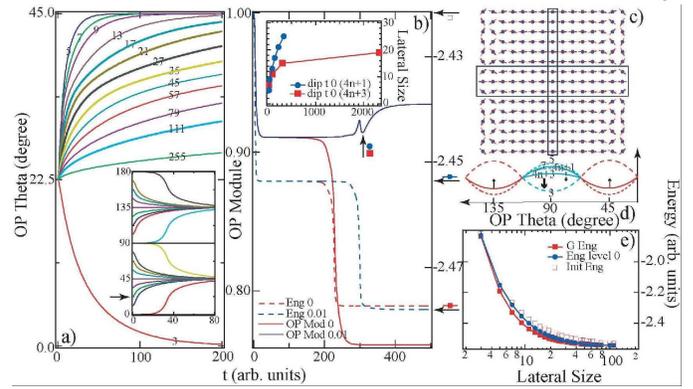}} \caption[Odd-L
dynamics] {(color) Spin dynamics for the odd-$L$ lattices. (a)
$L$-dependent evolution of the OP22.5 state ($L$ indicated on
curves). Inset: typical evolutions of various OP theta states at
intermediate $L$ ($L= 9$). (b) evolutions in unit energy and OP
module of OP0 and OP0.01 states ($L= 10$). The $L$-dependent
relaxation time position of the dip in the OP module curve of
OP0.01 is summarized in the inset. (c) the spin configuration ($L=
15$) corresponding to the level in the OP module curve of OP0
(OP0.01), namely the intermediate BD-OP0 in the text. Note that a
row dislocation and a column dislocation are as framed
respectively. (d) schematics showing the evolution of 3D free
energy surface in the transverse view. Numbers are $L$ values and
arrows indicate the relative motion of energy surface. (e) the
$L$-dependent energies of the ground state BD-OP45, the
intermediate BD-OP0 and the initial state OP theta (theta any)
(semilog).} \label{OddL}
\end{figure}

\begin{figure}[t!]
\vspace*{1cm} 
\centerline{\includegraphics[width=3.5in]{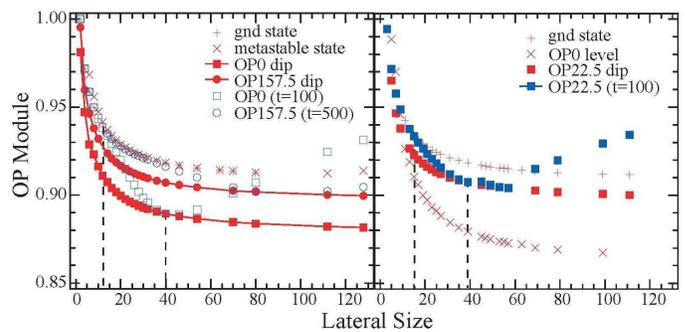}} \caption[OP
module summary] {(color) $L$-dependent OP modules of various
states and OP module samplings at given relaxation times for (a)
even-$L$, (b) odd-$L$. Note the vertical dash lines divide each of
the $t= 100$ curves into three characteristic regions.}
\label{OPmod}
\end{figure}

To obtain a systematic clarification, we simulate in Fig.
\ref{EvenL}(d) the transverse view of the imaginary free energy
surface\cite{FreeEngSurf}, which facilitates our understanding of
the OP rotation process following the boundary-distortion. The
convex free energy surface centering at OP theta $45^{\circ}$ is
depressed upon $L$ increase until its substitution by a concave at
$L= 6$. The two concaves at OP theta $45^{\circ}$ and
$135^{\circ}$ intercepts to form a watershed somewhere in between,
which is kept pushing asymptotically toward OP theta $90^{\circ}$
by the further lowering in energy of the OP theta $45^{\circ}$
concave and meanwhile enlarging its attraction area. The proximity
of the watershed induces some seemingly odd features in dynamics
of, for example, the OP22.5 at $L= 6$ (Fig. \ref{EvenL}(a)). As
summarized in the inset of Fig. \ref{EvenL}(b), there exists a
pronounced difference in boundary relaxation time for the two
initial states located symmetrically about OP theta $90^{\circ}$,
which tends to diminish as $L$ increases up to $18$. This
difference in time suggests the difference in physical relaxation
path taken respectively by the two initially symmetric states. The
latter is governed by the difference in topology of the free
energy surface due to the joint effect of the asymmetry in initial
state energy about OP90, which almost disappears at $L> 16$ (Fig.
\ref{EvenL}(e)), and the asymmetry by a finite displacement in OP
theta position of the watershed from $90^{\circ}$. The
$90^{\circ}$-axis OP symmetry recovery can also be evidenced by
the disappearance of the initial inequality in the final-state
$|OP|$ between BD-OP135 and BD-OP45 at roughly the same $L$ (Fig.
\ref{OPmod}(a)). From $L= 18$, the further modification on the
free energy surface topology occurs mainly along the energy axis
(vertically). As indicated by Fig. \ref{EvenL}(e), a gradual loss
of the transverse gradient precedes the final loss of longitudinal
gradient. The former is achieved by the flattening of both
concaves at the same time with the fully recovery of symmetry
about $45^{\circ}$-axis by a vertical alignment of the bottoms of
both concaves; the latter is expected when the initial and final
states converge in energy, which tends to flatten the OP theta
evolution curves (Fig. \ref{EvenL}(a)) and drives the boundary
relaxation time to be infinite (the inset of Fig. \ref{EvenL}(b)).

A similar outline is found for the story in the odd-$L$ case
though the details are different due to its difference in OP
symmetry. The free energy surface has an opposite topology between
$L= 3$ and $5$. The transformation in the OP theta evolution of
OP22.5 in Fig. \ref{OddL}(a) shows the appearance of a new
meta-stable state, BD-OP45. This only ground state for $L> 3$
takes over all the initial states, except the exact OP0 (inset of
Fig. \ref{OddL}(a)). Accordingly, in the $|OP|$ evolution, apart
from the BD states found corresponding to the module minimum as in
the even-$L$ case, there exists an intermediate flat region for
the OP0 initial states before the final arrival of a fully BD-OP0
state, which is metastable with a slightly higher energy and a
much reduced $|OP|$ than the ground state BD-OP45 (Fig.
\ref{OddL}(b)). A small deviation on the initial-state OP theta
from $0^{\circ}$ (OP0.01, in Fig. \ref{OddL}(b)) leads to an
ultimate fall onto BD-OP45 after a substantial stay in the
intermediate BD-OP0 state, as illustrated in Fig. \ref{OddL}(c).
This suggests the existence of a convex free energy surface
centering at OP theta $0^{\circ}$. On closer inspection, the spin
configuration of this intermediate state is composed of two
identical sub-lattices with even-$L$, which is formed by
introducing two topological dislocations, in row and column, into
the odd-$L$ lattice. To minimize the size of dislocation and
meanwhile to maximize the size of even-$L$ sub-lattice, there are
two different forms of the row dislocation depending on the
specific odd-$L$ value. For $L=4n+1$ ($n=1, 2,...$), an FM line
dislocation is formed; for $L=4n+3$, an elongated OP0 at $L= 3$ is
inserted as dislocation. The constructional difference leads to a
difference in the topology, and hence the flattening process, of
the convex free energy surface in response to the $L$ increase.
This is reflected by the difference in dynamics of this
intermediate state as in the inset of Fig. \ref{OddL}(b), where
its durations, quantified in the relaxation time position of the
dip in the OP0.01 dynamics, at various $L$'s are extracted and
found to follow two distinct paths.

Different from the even-$L$ case, the existence of the intrinsic
$90(0)^{\circ}$-axis OP symmetry simplifies the picture of the
$L$-dependent evolution of the free energy surface for the odd-$L$
lattice (Fig. \ref{OddL}(d)). A similar process of the global loss
of surface gradient is indicated by Fig. \ref{OddL}(e).
Interestingly, the $(4n+1)-(4n+3)$ difference doesn't show up in
the energy of the intermediate BD-OP0 state, reflecting again the
topological nature of this difference only for the proximity of OP
theta $0^{\circ}$. Though we can see clearly the different OP
symmetry governs the whole spin dynamics evolution on the two
kinds of lattices by means of the formation of a limited number of
meta-stable/intermediate states (with concave/convex free energy
surface, respectively) falling on the OP symmetry axes, the
similarity in large-$L$ evolution, combined with the gradual
recovery of a non-intrinsic $0(90)^{\circ}$-axis OP symmetry in
the even-$L$ case, point to the unification in the extreme
behavior at $L\rightarrow\infty$ when the even- and odd-$L$
lattice become practically indistinguishable.

However, if the system is allowed to relax onto its final state,
though through a sufficiently long time at large $L$ to complete
its boundary relaxation (insets of Fig. \ref{EvenL}(b) and
\ref{OddL}(b)), the difference in $|OP|$ of these BD states will
NOT be smeared out for both kinds of lattice as suggested by the
parallel saturations as seen for BD-OP135(45), BD-OP0 and
BD-OP157.5 on even-$L$ lattices (Fig. \ref{OPmod}(a)), BD-OP45,
BD-22.5 and intermediate BD-OP0 (Fig. \ref{OPmod}(b)).
Experimentally, the determination of the final state is restricted
by the accessible observation time. In Fig. \ref{OPmod}, examples
are shown for both kinds of lattices, a non-monotonous behavior is
expected in $|OP|$ sampled after a fixed relaxation time. For the
initial states other than those on the OP symmetry axes, three
well-defined regions are present depending on the comparison of
the sampling time with the boundary relaxation time and saturation
time. At the extreme case, where $L$ goes to infinite and the
even-odd difference naturally disappears, the boundary relaxation
of the given homogenous PBC state seems indiscernible within any
accessible experimental observation time, and thus, practically,
the PBC-predicted continuously degenerate state is regarded to be
stable.

This work was supported by the National Natural Science Foundation
of China, the Cheung Kong Program, the Hong Kong Qiushi Science
Foundation, and the Y. D. Fok Education Foundation.


%
%





%
%


\begin{thebibliography}{}
\bibitem{DipoleTermNote} Here $\Omega=(g\mu_B)^2$, $g$ is the Lande factor, $\mu_B$ the Bohr
magnetron.
\bibitem{OPsymmetryNote} For example, for the IP-AFM states, the equivalents for OP22.5
include OP67.5, OP202.5 and OP247.5 in even-L lattices, and
additionally include OP112.5, OP157.5, OP292.5 and OP337.5. The
equivalents share the same energy, OP module and all dynamical
details. For completeness, the denotations in text should include
the OP symmetry consideration.
\bibitem{FreeEngSurf} The imaginary 3D free energy surface is the result of the combined
 evolution of unit energy and OP theta. We named in text the OP
theta the x-axis (transverse), relaxation time the y-axis
(longitudinal).
\end{thebibliography}
\end{document}